\documentclass[
superscriptaddress,
preprint,
 amsmath,amssymb,
]{revtex4-1}

\usepackage{graphicx}
\usepackage{filecontents}
\usepackage{tabularx}
\usepackage{dcolumn}
\usepackage{bm}

\makeatletter
\let\@fnsymbol\@fnsymbol@latex
\@booleanfalse\altaffilletter@sw
\makeatother
    
\begin{document}

\preprint{APS/123-QED}

\title{Giant topological Hall effect in centrosymmetric tetragonal  Mn$_{2-x}$Zn$_x$Sb}

\author{Md Rafique Un Nabi}
\affiliation{Department of Physics, University of Arkansas, Fayetteville, AR 72701, USA}
\author{Aaron Wegner$^{\dagger}$}
\affiliation{Department of Physics, University of Arkansas, Fayetteville, AR 72701, USA}
\author{Fei Wang}
\affiliation{Department of Chemistry, Missouri State University, Springfield, Missouri, 65897, USA}
\author{Yanglin Zhu}
\affiliation{Department of Physics, Pennsylvania State University, University Park, Pennsylvania 16802, USA}
\author{Yingdong Guan}
\affiliation{Department of Physics, Pennsylvania State University, University Park, Pennsylvania 16802, USA} 
\author{Arash Fereidouni}
\affiliation{Department of Physics, University of Arkansas, Fayetteville, AR 72701, USA}
\author{Krishna Pandey}
\affiliation{Materials Science and Engineering Program, Institute for Nanoscience and Engineering, University of Arkansas, Fayetteville, AR 72701, USA}
\author{Rabindra Basnet}
\affiliation{Department of Physics, University of Arkansas, Fayetteville, AR 72701, USA}
\author{Gokul Acharya}
\affiliation{Department of Physics, University of Arkansas, Fayetteville, AR 72701, USA}
\author{Hugh O. H. Churchill}
\affiliation{Department of Physics, University of Arkansas, Fayetteville, AR 72701, USA}
\author{Zhiqiang Mao}
\affiliation{Department of Physics, Pennsylvania State University, University Park, Pennsylvania 16802, USA}
\author{Jin Hu$^{*}$}
\affiliation{Department of Physics, University of Arkansas, Fayetteville, AR 72701, USA}
\affiliation{Materials Science and Engineering Program, Institute for Nanoscience and Engineering, University of Arkansas, Fayetteville, AR 72701, USA}

\date{November 16, 2021}

\begin{abstract}

Topological magnetism typically appear in non-centrosymmetric compounds or compounds with geometric frustration. Here, we report the effective tuning of magnetism in centrosymmetric tetragonal Mn$_{2-x}$Zn$_{x}$Sb by Zn substitution. The magnetism is found to be closely coupled to the transport properties, giving rise to a very large topological hall effect with fine tuning of Zn content, which even persists to high temperature ($\sim$ 250 K). The further magnetoentropic analysis suggests that the topological hall effect is possibly associated with topological magnetism. Our finding suggests Mn$_{2-x}$Zn$_{x}$Sb is a candidate material for centrosymmetric tetragonal topological magnetic system, offers opportunities for studying and tuning spin textures and developing near room temperature spin-based devices.

\end{abstract}

\maketitle

\section{Introduction}

Magnetic materials with topological spin textures have attracted considerable attention for use in spintronic devices \cite{tokura}. Materials with magnetic skyrmions are considered useful for nonvolatile \cite{nonvolatile}, low power computing \cite{lowpower1, lowpower2, lowpower3} as topological magnetic materials are robust to perturbation due to the topological protection of the magnetic states. Magnetic phases with nontrivial topology have attracted attention for their ability to be manipulated by low current densities \cite{lowcurrent1, lowcurrent2, lowcurrent3, lowcurrent4} and with low dissipation \cite{dissipationless}. Nontrivial topological magnetic states most commonly arise from geometric frustration in hexagonal lattices \cite{frustration1, frustration2, frustration3, frustration4}, and spin-orbit coupling through the  Dzyaloshinskii-Moriya interaction \cite{DMI1, DMI2} such as chiral noncentrosymmetric B20 materials MnSi \cite{mnsi}, FeGe \cite{fege}, and Fe$_{1-x}$Co$_x$Si \cite{fecosi}. In contrast, in inversion symmetric  materials with no geometric frustration, topological spin textures are less common and normally require additional mechanisms, such as centrosymmetric tetragonal GdRu$_2$Si$_2$ which displays a skyrmion lattice owing to the interactions between itinerant electrons and magnetic order\cite{gdrusi}. 

Materials with layered crystal structure are of particular interest. These materials are ideal candidates for device applications as their structure may allow for exfoliation and easy integration into layered heterostructures, which opens a new route in exploring intrinsic magnetism in the 2D limit. To date, 2D magnetism has been observed in many hexagonal and trigonal materials, but remains elusive in other lattice types \cite{2d1, 2d2, 2d3}. Therefore, the discovery of tetragonal layered magnetic materials with tunable spin textures is highly desired for better understanding of low dimensional magnetism. One promising class of materials is Mn$_{2-x}$\textit{T}$_x$Sb, where \textit{T} is a transition metal such as Zn \cite{valkov, ZnDoped, GMR}, Cr \cite{CrDoped, CrDoped2}, Co \cite{CoDoped}, or Fe \cite{FeDoped}. As the parent compound of Mn$_{2-x}$\textit{T}$_x$Sb, Mn$_2$Sb crystallizes in a Cu$_2$Sb-type centrosymmetric tetragonal lattice with \textit{P}4/\textit{nmm} space group. It displays a ferrimagnetic (FI) order with spins collinear along the \textit{c}-axis below 550 K, and undergoes another magnetic transition with spins re-oriented within the \textit{ab} plane below 240 K  \cite{Mn2Sb}. The substitution of Mn by other transition metals \textit{T} tunes magnetic phases, which consequently modifies other properties such as cell volume \cite{valkov}, magnetostriction \cite{magnetostriction}, electronic transport \cite{valkov}, and the magnetocaloric effect \cite{MCE}. Such coupling of magnetism and physical properties, together with the layered structure and high electrical conductivity of these materials, makes Mn$_{2-x}$\textit{T}$_x$Sb a good material platform to study and tune the competing exchange interactions, as well as to develop new magnetic devices. 

Among transition metals, the non-magnetic Zn substitution to create Mn$_{2-x}$Zn$_x$Sb is particularly interesting. It has been reported that light Zn-substitution leads to a weak ferrimagnetic state at low temperatures \cite{valkov}. Further increasing Zn content to $x = 1$ can fully suppress the anti-parallel magnetic sublattice, leading to ferromagnetic (FM) order with the Curie temperature of 320 K in MnZnSb \cite{MnZnSb_Structure, MnZnSb}. Such room temperature ferromagnetism is of great interest for technological applications. As Mn$_{2-x}$Zn$_{x}$Sb has been ascribed to itinerant electron magnetism \cite{valkov} and displays tunable magnetic phases, it is an interesting candidate for nontrivial magnetic topology in a centrosymmetric tetragonal lattice. Under this motivation, we studied the magnetic and electronic properties of Mn$_{2-x}$Zn$_{x}$Sb compounds. We focused on Zn-rich samples (${x} > 0.5$) which has not been previously studied (except for ${x} = 1$). We observed rich magnetic phases in this material system and a very large topological Hall effect (THE) with a maximum $\rho_{xy}^T ~ \sim 2\  \mu\Omega$ occurring at high temperature up to 250 K. These observations, together with the magnetoentropic analysis, imply Mn$_{2-x}$Zn$_{x}$Sb can be a rare material platform for tunable topological magnetism in centrosymmetric tetragonal lattices.

\section{Experimental Methods}

Mn$_{2-x}$Zn$_{x}$Sb single crystals were grown by a self-flux method with Zn flux. Mn powder (99.6\%, Alpha Aesar), Zn powder (99.9\%, Alpha Aesar) and Sb powder (99.5\%, Beantown Chemical) with ratio 1+\textit{x}:6:1 were loaded in an alumina crucible inside evacuated fused quartz tubes. The tubes were heated to the maximum growth temperature over 30 hours, held for 3 days, and cooled down to 600$^\circ$C, followed by subsequent centrifuge to remove the excess flux. The maximum growth temperature varied with sample composition. For example, 800$^\circ$C for stoichiometric MnZnSb while 900$^\circ$C for other compositions. Millimeter-size crystals with metallic luster were obtained, as shown in the insets of Fig. 1b. Their compositions and crystallinity were checked by energy dispersive x-ray spectroscopy (EDS) and x-ray diffraction (XRD) (Fig. 1b), respectively. Throughout this work, the compositions shown below are measured compositions determined by EDS. The crystal structure were determined by single crystal XRD in a Bruker Apex I diffractometer with Mo \textit{K}$_\alpha$ radiation ($\lambda$ = 0.71073 Å). Resistivity and most of the magnetization measurements were carried out using a Quantum Design Physical Properties Measurement System. Some magnetization measurements were performed with a Quantum Design Superconducting Quantum Interference Device.

\section{Results}
As shown in Fig. 1a, the parent compound Mn$_{2}$Sb crystallizes in a layered tetragonal structure with two inequvalent Mn sites Mn(I) and Mn(II), whereas the Zn substitution has been found to occur at the Mn(II) site \cite{valkov, MnZnSb_Structure, GMR}. The systematic shift of the (00\textit{l}) XRD peaks with varying the Zn content indicates successful substitution in our samples, as shown in Fig. 1b. We have further performed crystal structure refinement using single crystal XRD on Mn$_{2-x}$Zn$_{x}$Sb, which have revealed that Zn-substitution maintains the crystal structure with the same space group \textit{P}4/\textit{nmm}, and the Zn substitution indeed occurs at the Mn(II) site. In Supplemental Materials \cite{SM} we provide the structure parameters resolved using JANA2006 \cite{JANA2006}. 

\begin{figure}[t!]
\includegraphics[width=\textwidth]{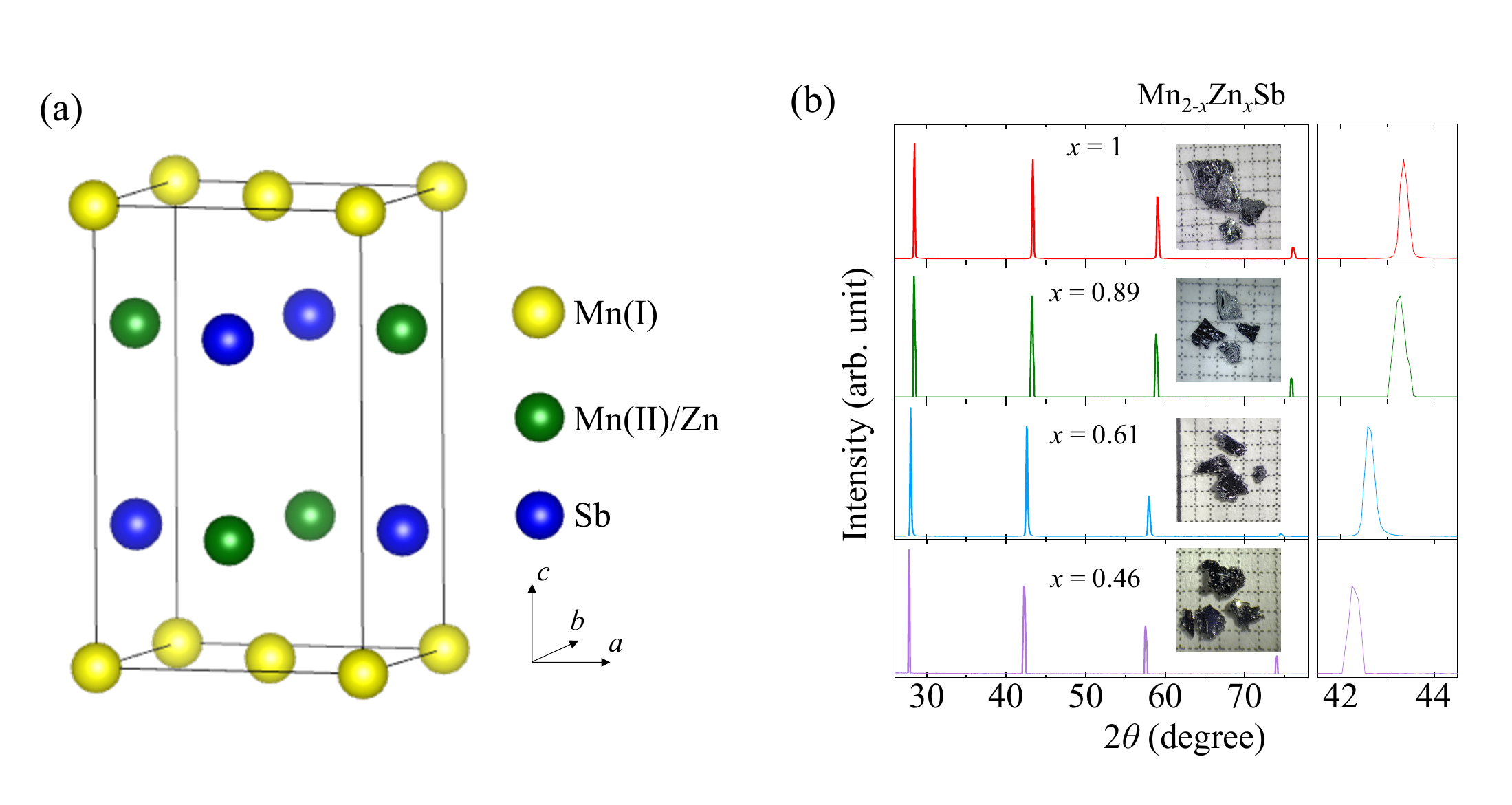}
\caption{\label{Figure 1} (a) Crystal structure for Mn$_{2-x}$Zn$_{x}$Sb. (b) Left: Images of a few representative Mn$_{2-x}$Zn$_{x}$Sb single crystals used in this work and their (00\textit{L}) series x-ray diffraction patterns. Right: zoom of the XRD pattern to show the systematic shift of the (003) diffraction peak.}
\end{figure}

In Mn$_{2}$Sb, Mn(I) and Mn(II) sublattices carry opposite magnetic moments which leads to two FI phases. The high temperature FI phase (denoted as HT-FI) occurs between 240 and 550K, which is made up of inequivalent Mn(I) and Mn(II) magnetic sublattices with moments oriented out-of-plane. Below 240 K, the moments of both magnetic sublattices are re-oriented to in-plane direction which forms the low temperature FI phase (denoted as LT-FI) \cite{Mn2Sb}. The schematic magnetic structures are  shown in the insets of Fig. 2c. Therefore, the dilution of Mn(II) magnetism by non-magnetic Zn substitution can significantly affect the magnetism. The magnetic transition temperatures have been reported to be suppressed by light Zn-substitution ($x < 0.3$) \cite{valkov, ZnDoped, GMR}. In this work, we clarified the magnetism of the heavily Zn-substituted compounds. Upon increasing the Zn content, the out-of-plane (\textbf{H}//\textit{c}, Fig. 2a) and in-plane (\textbf{H}//\textit{c}, Fig. 2b) magnetic susceptibility measurements indicate that the two transition temperatures systematically decrease. The HT-FI ordering temperature  saturates to 300 - 310 K. For the LT-FI phase, the order is completely suppressed when Mn(II) is fully replaced by Zn (${x} = 1$), leaving a room temperature ferromagnetism with $T_c\approx$ 310 K that corresponds to the magnetic ordering of the Mn(I) sublattice. These observations indicate that the magnetic phases in Mn$_{2-x}$Zn$_{x}$Sb are determined by the coupling of Mn(I) and Mn(II) magnetic lattices.

\begin{figure}[t!]
\includegraphics[width=\textwidth]{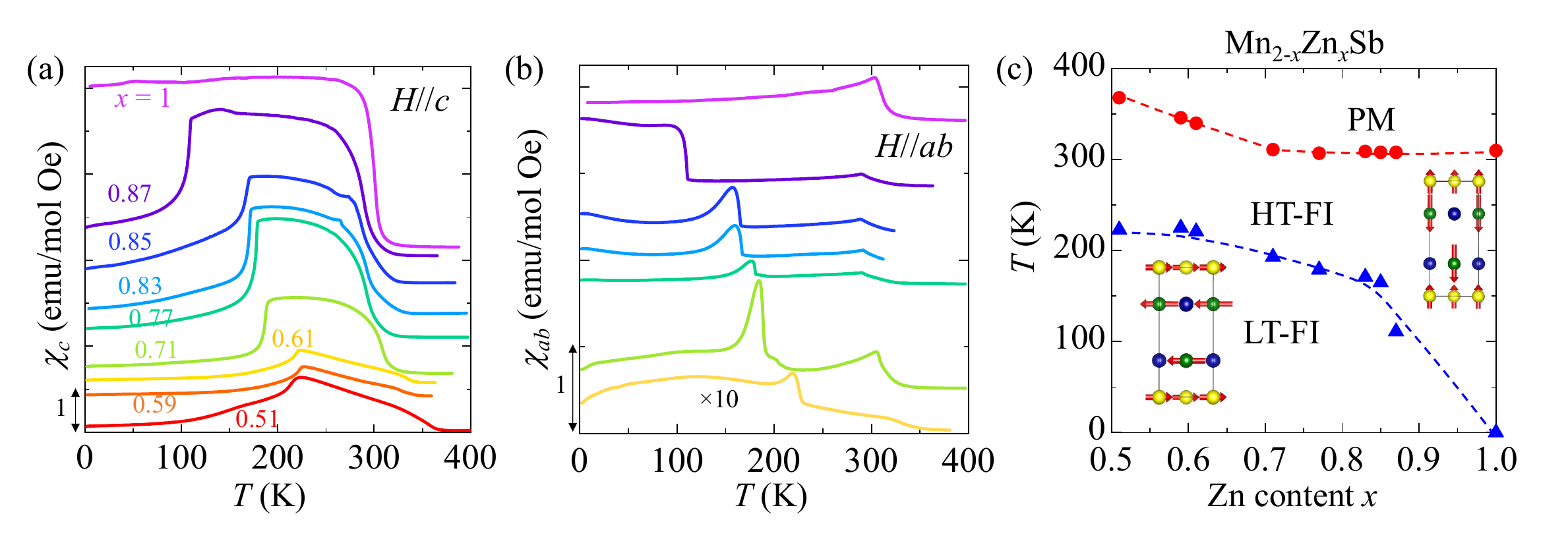}
\caption{\label{Figure 2} (a-b) Temperature dependence of magnetic susceptibility for Mn$_{2-x}$Zn$_{x}$Sb measured with magnetic field of 0.1 T applied (a) along the \textit{c}-axis ($H\parallel c$) and (b) within the \textit{ab}-plane ($H\parallel ab$). Data for each composition are shifted for clarity. The same color code is used for (a) and (b). (c) Magnetic phase diagram constructed from the susceptibility measurements presented in (a) and (b). Inset: the schematic magnetic structures for the HT-FI and LT-FI phases.}
\end{figure}

The evolution of magnetism in heavily Zn-substituted ($x > 0.5$) samples is summarized by a magnetic phase diagram in Fig. 2c, together with the possible magnetic structures for HT-FI and LT-FI phases adopted from the earlier studies \cite{Mn2Sb, valkov}. For MnZnSb which is FM, the magnetic structure is similar to that for the HT-FI phase, except that the Mn(II) sublattice is replaced by non-magnetic Zn \cite{MnZnSb}. Though the reported magnetic structures for HT-FI and LT-FI phases were originally determined for lightly Zn-substituted samples \cite{valkov}, they agrees well with the observed magnetic properties of our Zn-rich samples. As shown in Fig. 2a, when {H}//{c}, susceptibility $\chi_c$ is the greatest in the HT-FI phase but drops steeply upon entering the LT-FI phase. In contrast, under in-plane field {H}//{ab}, the LT-FI phase susceptibility $\chi_{ab}$ is comparable or greater than that of the HT-FI phase. These results are consistent with the proposed magnetic structures with the easy axis being out-of-plane for the HT-FI phase but in-plane for the LT-FI phase (Fig. 2c, insets), which has also been observed in Co-doped Mn$_{2}$Sb \cite{CoDoped}.

\begin{figure}[t!]
\includegraphics[width=\textwidth]{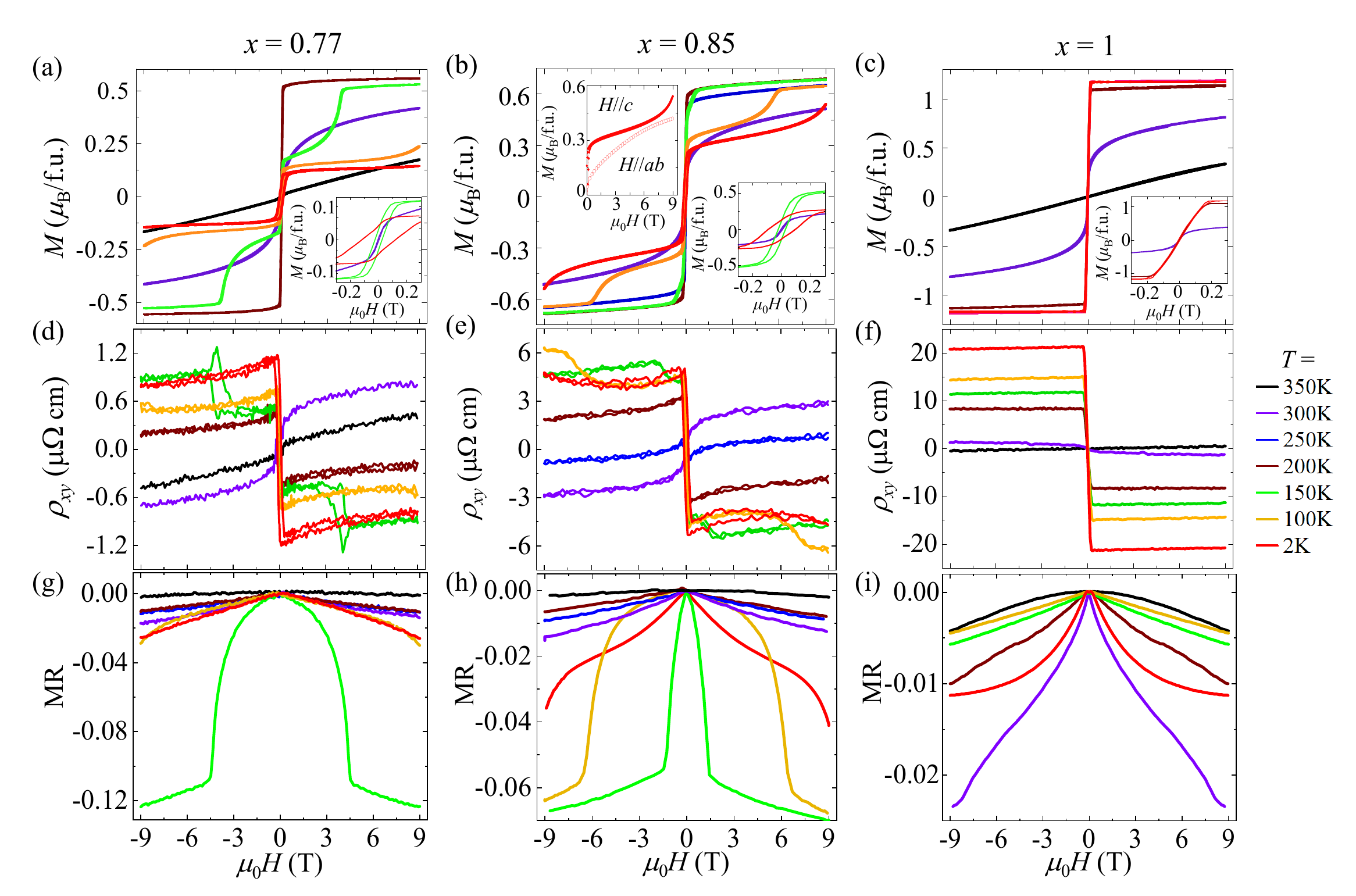}
\caption{\label{Figure 3} Magnetic field dependence of (a-c) isothermal magnetization $M$, (d-f) Hall resistivity $\rho_{xy}$, and  (g-i) normalized magnetoresistivity, MR = [$\rho_{xx}(H)$ - $\rho_{xx}(H=0)$]/$\rho_{xx}(H=0)$ for $x = 0.77$, $x = 0.85$, and $x = 1$ samples, measured with \textbf{H}//\textit{c}. Metamagnetic transitions below the HT-FI transition temperatures are manifested by the jump at higher fields at 150 K in $x = 0.77$ and 100 K in $x = 0.85$ sample. For each of those three compositions, identical samples were used for Hall effect and magnetoresistivity measurements. Lower insets in (a-c): zoom-in of the low field magnetization. Upper inset in (b): out-of-plane (\textbf{H}//\textit{c}) and in-plane (\textbf{H}//\textit{ab}) magnetization for the $x = 0.85$ sample at $T = 2$ K. The same color code is used for all panels.}
\end{figure}

Magnetic properties of Zn-substituted samples have also been characterized by isothermal field-dependent magnetization measurements. In Figs. 3a-c we present the field dependent magnetization $M(H)$ for a few representative samples $x = 0.77, 0.85$, and $1$, measured with \textbf{H}//\textit{c}. The low field hysteresis loop and the high field moment saturation behavior are consistent with the nature of ferrimagnetism. In the $x = 0.77$ and $0.85$ samples with mixed Mn(II)/Zn plane, a strong metamagnetic transition can be observed below the LT-FI transition temperature (e.g., 150 K for the $x = 0.77$ sample and 100 K for the $x = 0.85$ sample), which shifts to higher fields with lowering temperature and is absent in  measurements under in-plane magnetic field ({H}//{ab}, see Fig. 3b, inset). Those observations imply a field-driven spin canting and reorientation from the LT-FI magnetic structure to the HT-FI structure (see magnetic structures in Fig. 2c). In MnZnSb ($x=1$), the full replacement of the Mn(II) lattice by Zn plane leads to a typical FM behavior with saturation moment around 1.2 $\mu_B$/Mn, consistent with earlier studies \cite{MnZnSb}. In addition, as shown in the insert of Fig. 3c, the width of the hysteresis loop is negligible in MnZnSb. The small magnetic coercivity further indicates soft room temperature ferromagnetism.

The highly tunable magnetic phases in Mn$_{2-x}$Zn$_{x}$Sb provides an good opportunity to study the evolution of electronic transport properties with fine-tuning of magnetism. The strong coupling of magnetism and electron transport in this material system has been revealed by the observation of strong anomalous Hall effect (AHE) in all Mn$_{2-x}$Zn$_{x}$Sb samples we have measured. Figures. 3d-f present the field dependence of Hall resistivity $\rho_{xy}(H)$ for a few representative samples with $x = 0.77, 0.85$, and 1. The jumps in Hall resistivity $\rho_{xy}$ match well with the magnetization jumps, both near the zero field and at the metamagnetic transition (Figs. 3a-3c). In addition to the low field hysteresis loop that was also seen in the magnetization measurements, the $x = 0.77$ sample also display a hysteresis loop in the field scan at the metamagnetic transition around 4 T at 150 K (Fig. 3d), which is not observed in the magnetization measurement (Fig. 3a). In addition, as shown in Figs. 3g-3i, the magnetotransport measurements reveal negative magnetoresistivity (MR) which is consistent with the nature of ferrimagnetism, as well as the clear kinks at magnetization transitions.

One interesting feature in our transport study is the presence of THE in Mn$_{2-x}$Zn$_{x}$Sb in $\rho_{xy}$ data. In materials with nontrivial spin texture, conduction electrons acquire a Berry phase that gives rise to a fictitious magnetic field, leading to additional THE contribution to Hall resistivity, as given by \cite{THEformula}:

\begin{equation}
\rho_{xy} =  R_0B + S_H \rho_{xx}(B,T)^2M(B,T) + \rho_{xy}^T
\end{equation}

where $R_0 = \frac1{nq}$ is the ordinary Hall coefficient, $S_H$ is the anomalous Hall coefficient, $\rho_{xx}$ is the longitudinal resistivity, and $M$ is the magnetization. The first, second, and third terms in Eq. 1 correspond to ordinary Hall effect (OHE), anomalous Hall effect (AHE), and topological Hall effect (THE). For the AHE component, we choose the widely used quadratic dependence for longitudinal resistivity (i.e., AHE $\propto\rho_{xx}(B,T)^2$), which is applicable for extrinsic AHE in good metals ($\sigma \sim 10^4 - 10^6$ ($\Omega$ cm)$^-1$), or AHE with an intrinsic Berry phase mechanism \cite{AHEreview,THEformula}. Although extrinsic AHE originating from a skew scattering mechanism can show a linear dependence (i.e., AHE $\propto\rho_{xx}$) \cite{AHEreview,THEformula}, it cannot fit our measured data. For the THE component $\rho_{xy}^T$, it can arise due to topological spin texture in the real space, or the Berry phase in momentum space through a topological band structure. 

For FM MnZnSb in which the Mn(II) plane is fully replaced by Zn, $\rho_{xy}$ can be well reproduced by ordinary and anomalous Hall components with a negative anomalous Hall coefficient $S_H$. However, Mn$_{2-x}$Zn$_{x}$Sb samples with mixed Mn(II)/Zn lattice plane display distinct features. First, $S_H$ reverses its sign from positive to negative with lowering the temperature (Figs. 3d and 3e). The mechanism for such sign change is unclear, but appears to be associated with changing magnetic easy axis from out-of-plane in HT-FI phase to in-plane in LT-FI phase (see magnetic structures in Fig. 2b). Second, $\rho_{xy}$ at low fields does not scale to magnetization, indicating the presence of a THE component $\rho_{xy}^T$, which can be extracted by subtracting the ordinary and anomalous Hall contributions from the total measured $\rho_{xy}$, from the fits using the above Eq. 1. A typical example showing the extraction of $\rho_{xy}^T$ at $T=2$ K for the $x= 0.85$ sample is shown in Fig. 4a, in which the measured magnetic field dependence of Hall resistivity has been decomposed to the OHE, AHE, and the remaining THE component. The extracted $\rho_{xy}^T(H)$ exhibits a peak near the zero field, which is gradually suppressed upon increasing magnetic field. Such a field dependence for $\rho_{xy}^T(H)$ is reminiscent of the THE in skyrmion systems. In addition to magnetic field, increasing temperature also suppresses $\rho_{xy}^T$, which can be directly seen from the absence of the low field feature in the measured Hall effect data (Figs. 3d and 3e). Similar behavior has also been observed in all other Mn$_{2-x}$Zn$_{x}$Sb samples with varied $\rho_{xy}^T$ amplitude, except for the $x = 1$ sample. In Fig. 4b we summarized the extracted maximum $\rho_{xy}^T$ in color map to present the temperature and composition dependence of the THE in Mn$_{2-x}$Zn$_{x}$Sb. It can be clearly seen that THE occurs with Zn substitution for Mn(II), becomes the strongest in the $x= 0.85$ sample with a large $\rho_{xy}^T \sim 2\  \mu\Omega$ , and diminishes when Mn(II) is fully replaced by Zn. Interestingly, the sizable THE even persists at high temperatures (250 K), which is large compared to other materials showing THE close to room temperature \cite{HT_THE1, HT_THE2, HT_THE3, HT_THE4, HT_THE5, HT_THE6, HT_THE7}.

\begin{figure}[t!]
\includegraphics[width=\textwidth]{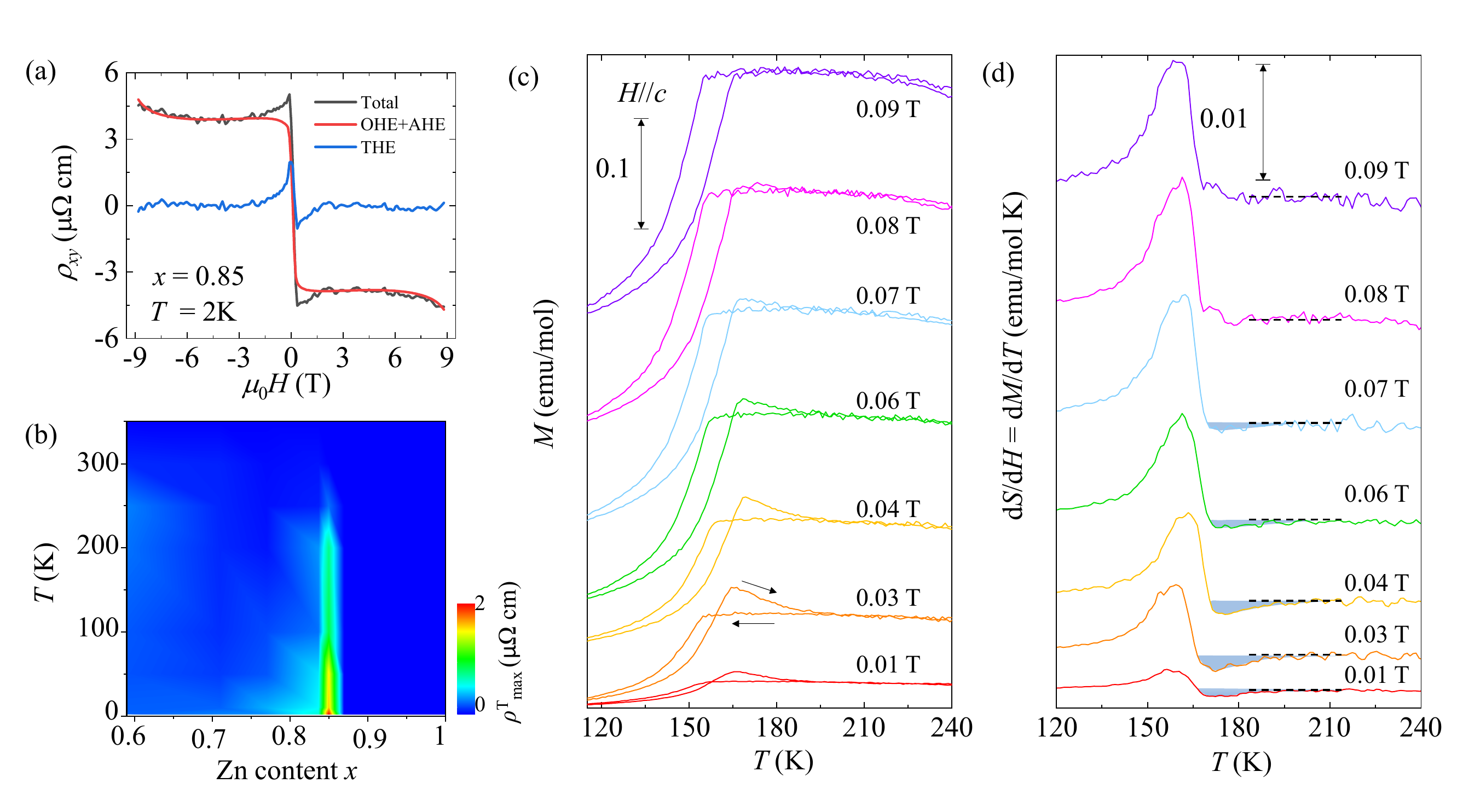}
\caption{\label{Figure 4} (a) Separation of the ordinary (OHE), anomalous (AHE), and the topological (THE) Hall components for the $x = 0.85$ sample at $T = 2$ K according to Eq. 1. (b) Color map for the maximum topological Hall resistivity for various Mn$_{2-x}$Zn$_{x}$Sb samples and at various temperatures. (c) Low field ($< 0.1$ T) susceptibility for the $x = 0.85$ sample, measured with both cooling and heating as indicated by the arrows. Data collected at various fields are shifted for clarity. (d) $(dS/dH)_T=(dM/dT)_H$ for the $x = 0.85$ sample, obtained from (c). Data collected at various fields are shifted for clarity, with the dashed lines indicating the zero value.}
\end{figure}

In magnetic materials, large THE is expected \cite{largeTHE} and observed \cite{THEnoncollinear, THEnoncollinear2} with noncollinear antiferromagnetism, which has been attributed to the Berry curvature-driven intrinsic anomalous Hall effect \cite{largeTHE}. As a specific scenario, the formation of skyrmion lattice with topological spin texture is known to cause large THE \cite{frustration4}. In addition, there is also a proposal of artificial THE caused by signals from multiple magnetic phases in inhomogeneous or polycrystalline samples \cite{CrTe,CoGd}, which can be excluded because no secondary phases have been probed in structural and magnetic property characterizations on our single crystalline samples. To clarify the mechanism of the large THE, we have performed the temperature dependent magnetization for the $x= 0.85$ sample under several perpendicularly applied fields (\textbf{H}//\textit{c}), as shown in Fig. 4c. Here we focus on low field ($< 0.1$ T) measurements because THE diminishes at high fields (Fig. 4a). Here the data were taken during both cooling and heating, in contrast to the conventional magnetization measurements in which the data are usually collected with increasing temperature. Our measurements reveal a field-dependent ordering temperature which shifts to lower values with increasing field. Furthermore, there is a thermal hysteresis in the HT-FI to LT-FI transition, and an anomaly peak immediately above the transition. The anomaly peak only occurs upon heating from the LT-FI phase, which is suppressed at higher fields ($> 0.07$ T). This magnetization anomaly, which occurs as a precursor to the magnetic transition, implies that there is an intermediate phase depending on the temperature history of the sample. Such an anomaly peak in magnetization has also been observed in other materials with non-trivial spin textures \cite{FeGeSkyrmion, CoZnMn}.

The transitions between topologically distinct spin states in real space are characterized by entropy changes which can be conveniently revealed by temperature dependent magnetization measurements via the Maxwell relation $(dS/dH)_T=(dM/dT)_H$ \cite{FeGeSkyrmion}. The derivative of the warming data of magnetization is presented in Fig. 4d, which displays a positive peak corresponding to the magnetic transition in $M(T)$. At low fields, a negative peak (the blue shaded region) in $dS/dH$ occurs immediately above the magnetic transition temperature and extend to ~220 K, indicating a reduced magnetic entropy. This negative peak disappears at higher field ($> 0.07$ T). The temperature and field dependence of the negative peak agrees well with the emergence of THE in the $x= 0.85$ sample (Figs. 4a and 4b). Similar magnetoentropy signatures have also been observed in FeGe \cite{FeGeSkyrmion} and Co$_x$Zn$_y$Mn$_z$ \cite{CoZnMn} which are attributed to the formation of skyrmion lattice. Though direct probes such as Lorentz transmission electron microscopy or small-angle neutron scattering are needed to clarify the existence of skyrmion phase in Mn$_{2-x}$Zn$_{x}$Sb, our observations indeed imply the development of a distinct spin state prior to the HT-FI to LT-FI phase transition.

\section{Discussion and Conclusion}
The magnetoentropic analysis indicates the presence of an intermediate phase that occurs only at low field, which further suggests that the THE is due to a real space Berry phase caused by a topological magnetic spin texture. One possibility is the skyrmion lattice phase stabilized by magnetic fluctuations near the phase boundary between the HT-FI and LT-FI phases. Centrosymmetric material possessing magnetic skyrmion or skyrmion-like features have been discovered such as SrFe$_{1-x}$Co$_x$O$_3$ and MnNiGa \cite{SFCO,MnNiGa}. Large THE in tetragonal lattice has also been reported in Mn$_2$PtSn which is non-centrosymmetric \cite{Mn2PtSn}. However, a skyrmion lattice phase is rare in the lattice which is centrosymmetric and tetragonal. In centrosymmetric tetragonal Mn$_{2-x}$Zn$_{x}$Sb, magnetic fluctuations and frustrations may be introduced by partial replacement of the magnetic sublattice. Furthermore, It has been reported that coupling between itinerant electron states and localized magnetic moments can lead to a skyrmion lattice without geometric frustration. These exchange interactions have been shown to lead to a square skyrmion lattice when stabilized by magnetic anisotropy in materials such as GdRu$_2$Si$_2$ through both experimental observations \cite{gdrusi, nanometric} and Monte Carlo simulations \cite{gdrusi, squareskyrmion}. 
More direct probes are needed to clarify the spin texture in Mn$_{2-x}$Zn$_{x}$Sb. Providing large THE in centrosymmetric tetragonal lattice is rare \cite{EuAl4}, the large THE occurring at very high temperatures and its high tunablity with compositions indicate that magnetically disordered Mn$_{2-x}$Zn$_{x}$Sb is a good candidate material for a tunable topological magnetism.

\textbf{Acknowledgement}

This work is primarily supported by the U.S. Department of Energy, Office of Science, Basic Energy Sciences program under Award No. DE-SC0019467 (support for personnel, sample synthesis, major transport and magnetization measurements, and data analysis). H.C. acknowledges NSF award DMR-1848281 (support for part of transport measurements). R.B. acknowledges the support from Chancellor’s Innovation and Collaboration Fund at the University of Arkansas (support for personnel). Z,Q.M. acknowledges the support by the US National Science
Foundation under grants DMR 1917579 and 1832031.

$^{*}$ jinhu@uark.edu
$^{\dagger}$ wegner@uark.edu


\begin{thebibliography}{99}

\bibitem{tokura} Y. Tokura, K. Yasuda, and A. Tsukazaki. \textit{Nature Reviews Physics} 1, 126 (2019).

\bibitem{nonvolatile}W. Kang, Y. Huang, X. Zhang, Y. Zhou, and W. Zhao.  \textit{Proceedings of the IEEE} 104, 2040, (2016).

\bibitem{lowpower1}Z. Zhang, Y. Zhu, Y. Zhang, K. Zhang, J. Nan, Z. Zheng, Y. Zhang, and W. Zhao. \textit{IEEE Electron Device Letters} 40, 1984 (2019).

\bibitem{lowpower2} R. Tomasello, E. Martinez, R. Zivieri, L. Torres, M. Carpentieri, and G. Finocchio. \textit{Scientific Reports} 4, 6784. (2014).

\bibitem{lowpower3} Y. Liu, N. Lei, C. Wang, X. Zhang, W. Kang, D. Zhu, Y. Zhou, X. Liu, Y. Zhang, and W. Zhao. \textit{Physical Review Applied} 11, 014004 (2019).

\bibitem{lowcurrent1}X. Z. Yu, N. Kanazawa, W. Z. Zhang, T. Nagai, T. Hara, K. Kimoto, Y. Matsui, Y. Onose, and Y. Tokura. \textit{Nature Communications} 3, 988 (2012).

\bibitem{lowcurrent2} W. Jiang, P. Upadhyaya, W. Zhang, G. Yu, M.B. Jungfleisch, F. Y. Fradin, J. E. Pearson, Y. Tserkovnyak, K.L. Wang, O. Heinonen, and S.G. Te Velthuis. \textit{Science} 349, 283 (2015).

\bibitem{lowcurrent3} S. Woo, K. Litzius, B. Krüger, M. Im, L. Caretta, K. Richter, M. Mann, A. Krone, R.M. Reeve, M. Weigand, and P. Agrawal. \textit{Nature materials} 15, 501, (2016).

\bibitem{lowcurrent4} K. Litzius, I. Lemesh, B. Krüger, P. Bassirian, L. Caretta, K. Richter, F. Büttner, K. Sato, O.A. Tretiakov, J. Förster, R.M. Reeve.  \textit{Nature Physics} 13, 170 (2017).

\bibitem{dissipationless} F. Jonietz, S. Mühlbauer, C. Pfleiderer, A. Neubauer, W. Münzer, A. Bauer, T. Adams, R. Georgii, P Böni, R.A. Duine, and K. Everschor. \textit{Science} 330, 1648 (2010).

\bibitem{frustration1}T. Okubo, S. Chung, and H. Kawamura. \textit{Physical Review Letters} 108, 017206 (2012).

\bibitem{frustration2}P. A. Kotsanidis, J. K. Yakinthos, and E. Gamari-Seale. \textit{Journal of magnetism and magnetic materials} 87, 199 (1990).

\bibitem{frustration3} J-H. Kim, A. Jain, M. Reehuis, G. Khaliullin, D. C. Peets, C. Ulrich, J. T. Park, E. Faulhaber, A. Hoser, H.C. Walker, and D.T. Adroja. \textit{Physical Review Letters} 113, 147206 (2014).

\bibitem{frustration4} K. Takashi, T. Nakajima, M. Hirschberger, A. Kikkawa, Y. Yamasaki, H. Sagayama, H. Nakao, Y. Taguchi, T. Arima, and Y. Tokura.  \textit{Science} 365, 914 (2019).

\bibitem{DMI1} I. Dzyaloshinsky. \textit{Journal of Physics and Chemistry of Solids} 4, 241 (1958). 

\bibitem{DMI2} T. Moriya. \textit{Physical Review} 120, 91 (1960).

\bibitem{mnsi} C. Pappas, E. Lelievre-Berna, P. Falus, P. M. Bentley, E. Moskvin, S. Grigoriev, P. Fouquet, and B. Farago. \textit{Physical Review Letters} 102, 197202 (2009).

\bibitem{fege} X. Z. Yu, N. Kanazawa, Y. Onose, K. Kimoto, W. Z. Zhang, S. Ishiwata, Y. Matsui, and Y. Tokura. \textit{Nature Materials} 10, 106, (2011).

\bibitem{fecosi} W. Münzer, A. Neubauer, T. Adams, S. Mühlbauer, C. Franz, F. Jonietz, R. Georgii, P. Böni, B. Pedersen, M. Schmidt, and A. Rosch.  \textit{Physical Review B} 81, 041203 (2010).

\bibitem{gdrusi}Y. Yasui, C.J. Butler, N.D. Khanh, S. Hayami, T. Nomoto, T. Hanaguri, Y. Motome, R. Arita, T. Arima, Y Tokura, and S. Seki. \textit{Nature Communications}11, 5925 (2020).

\bibitem{2d1}B. Huang, G. Clark, E. Navarro-Moratalla, D. Klein, R. Cheng, K. Seyler, D. Zhong, E. Schmidgall, M. McGuire, D. Coden, and W. Yao. \textit{Nature} 546, 270 (2017).

\bibitem{2d2}C. Gong, L. Li, Z. Li, H. Ji, A. Stern, Y. Xia, T. Cao, W. Bao, C. Wang, Y. Wang, and Z.Q. Qiu. \textit{Nature} 546, 265 (2017).

\bibitem{2d3}Z. Fei, B. Huang, P. Malinowski, W. Wang, T. Song, J. Sanchez, W. Yao, D. Xiao, X. Zhue, A.F. May, and W. Wu. \textit{Nature materials} 17, 778 (2018).

\bibitem{valkov} V. Val’Kov, A. Golovchan, I. Bribanov, V. Kamenev, O. Iesenchuk, A. Sivachenko, and N. Kabdin. \textit{Low Temperature Physics} 33, 70 (2007).

\bibitem{ZnDoped}V. Ryzhkovskii, and V. Mitsiuk. \textit{Inorganic Materials} 46, 581 (2010).

\bibitem{GMR}Q. Zhang, Y. Zhang, Y. Li, J. Du, W. Feng, D. Li, and Z. Zhang. \textit{Journal of Physics D: Applied Physics} 41, 095007 (2008).

\bibitem{CrDoped} A. Austin, E. Adelson, and W. H. Cloud. \textit{Physical Review} 131, 1511, (1963).

\bibitem{CrDoped2} K. Nakagawa, Y. Miyazaki, N. Mitsuishi, M. Sakano, T. Yokouchi, K. Ishizaka, and Y. Shiomi. \textit{Journal of the Physical Society of Japan} 89, 124601 (2020).

\bibitem{CoDoped} J. Wilden, A. Hoser, M. Chikovani, J. Perßon, J. Voigt, K. Friese, and A. Grzechnik. \textit{Inorganics} 6, 113 (2018).

\bibitem{FeDoped} S. Funahashi. \textit{Journal of magnetism and magnetic materials} 31, 595-596, (1983).

\bibitem{Mn2Sb} M. Wilkinson, N. Gingrich, and C. Shull. \textit{Journal of Physics and Chemistry of Solids} 2, 289 (1957).

\bibitem{magnetostriction}M. Bartashevich, T. Goto, N. Baranov, and V. Gaviko. \textit{Physica B: Condensed Matter} 351, 71 (2004).

\bibitem{MCE} N.Y. Pankratov, V.I. Mitsiuk, V.M. Ryzhkovskii, and S.A. Nikitin.  \textit{Journal of Magnetism and Magnetic Materials} 470, 46 (2019).

\bibitem{MnZnSb_Structure}V. Johnson, and W. Jeitschko. \textit{Journal of Solid State Chemistry} 22, 71 (1977).

\bibitem{SM} See Supplemental Material at (link provided by publisher)

\bibitem{JANA2006} V. Petricek , M. Dusek, and L. Palatinus. \textit{Zeitschrift Kristallographie} 229, 345 (2014).

\bibitem{MnZnSb} H. Matsuzaki, S. Endo, Y. Notsu, F. Ono, T. Kanomata, and T. Kaneko. \textit{Japanese Journal of Applied Physics} 32, 271 (1993).

\bibitem{THEformula} P. Swekis, A. Markou, D. Kriegner, J. Gayles, R. Schlitz, W. Schnelle, S. Goennenwein, and C. Felser.  \textit{Physical Review Materials} 3, 013001 (2019).

\bibitem{AHEreview} N. Nagaosa, J. Sinova, S. Onoda, A. H. MacDonald, and N. P. Ong, \textit{Rev. Mod. Phys.} 82, 1539 (2010).

\bibitem{HT_THE1}X. Li, C. Collignon, L. Xu, H. Zuo, A. Cavanna, U. Gennser, D. Mailly, B. Fauqué, L. Balents, Z. Zhu, and K. Behnia. \textit{Nat. Comm.} 10, 3021 (2019).

\bibitem{HT_THE2}X. Xiao, L. Peng, X. Zhao, Y. Zhang, Y. Dai, J. Guo, M. Tong, J. Li, B. Li, W. Liu, J. Cai, B. Shen, and Z. Zhang. \textit{Appl. Phys. Lett.} 114, 142404 (2019).

\bibitem{HT_THE3}S. Sen, C. Singh, P. K. Mukharjee, R. Nath, and A. K. Nayak. \textit{Phys. Rev. B} 99, 134404
(2019).

\bibitem{HT_THE4}Y. Li, B. Ding, X. Wang, H. Zhang, W. Wang, and Z. Liu. \textit{Appl. Phys. Lett.} 113, 062406
(2018).

\bibitem{HT_THE5}J.C. Gallagher, K. Y. Meng, J. T. Brangham, H.L. Wang, B. D. Esser, D. W. McComb,
11 and F. Y. Yang. \textit{Phys. Rev. Lett.} 118, 027201(2017).

\bibitem{HT_THE6}X. Zheng, X. Zhao, J. Qi, X. Luo, S. Ma, C. Chen, H. Zeng, G. Yu, N. Fang, S.U> Rehman, and W. Ren. \textit{Applied Physics Letters} 118, 072402 (2021).

\bibitem{HT_THE7}M. Leroux, M.J. Stolt, S. Jin, D.V. Pete, C. Reichhardt, and B. Maiorov. \textit{Scientific reports}, 8, 15510 (2018).


\bibitem{largeTHE} H. Chen, Q. Niu, and A. H. MacDonald.  \textit{Physical Review Letters} 112. 017205 (2014).

\bibitem{THEnoncollinear} C. Sürgers, G. Fischer, P. Winkel, and H. Löhneysen. \textit{Nature communications} 5, 3400 (2014).

\bibitem{THEnoncollinear2} J. Taylor, A. Markau, E. Lesne, P.K. Sivakumar, C. Luo, F. Radu, P. Werner, C. Felser, and S. Parkin. \textit{Physical Review B} 101, 094404 (2020).

\bibitem{CrTe} Y. He, J. Kroder, J. Gayles, C. Fu, Y. Pan, W. Schnelle, C. Felser, and G. H. Fecher. \textit{Appl. Phys. Lett.}, 117, 052409 (2020).

\bibitem{CoGd} T. Fu, S. Li, X. Feng, Y. Cui, J. Yao, B. Wang, J. Cao, Z. Shi, D. Xue, and X. Fan. \textit{Phys. Rev. B}, 103, 064432 (2021).

\bibitem{FeGeSkyrmion}J. Bocarsly, R. Need, R. Seshadri, and S. Wilson. \textit{Physical Review B} 97, 100404 (2018).

\bibitem{CoZnMn} J. Bocarsly, C. Heikes, C. Brown, S. Wilson, and R. Seshadri. \textit{Physical Review Materials}, 3, 014402 (2019).

\bibitem{SFCO} S. Chakraverty, T. Matsuda, H. Wadati, J. Okamoto, Y. Yamasaki, H. Nakao, Y. Murakami, S. Ishiwata, M. Kawasaki, Y. Taguchi, Y. Tokura, and H. Y. Hwang. \textit{Phys. Rev. B}, 88, 220405(R) (2013)

\bibitem{MnNiGa} W. Wang, Y. Zhang, G. Xu, L. Peng, B. Ding, Y. Wang, Z. Hou, M. Zhang, X. Li, E. Liu, S. Wang, J. Cai, F. Wang, J. Li, F. Hu, G. Wu, B. Shen, X.-X. Zhang. \textit{Adv. Mat.}, 24, 6887 (2016)

\bibitem{Mn2PtSn} Z.H. Liu, A. Burigu, Y.J. Zhang, H. M. Jafria, X.Q. Ma, E.K. Liu, W.H. Wang, G.H. Wu. \textit{Scripta Materialia}, 143, 122 (2018)



\bibitem{nanometric}N.D. Khanh, T. Nakajima, X. Yu, S. Gao, K. Shibata, M. Hirschberger, Y. Yamasaki, H. Sagayama, H. Nakao, L. Peng, and K. Nakajima. \textit{Nature Nanotechnology} 15, 444 (2020)

\bibitem{squareskyrmion}S. Hayami, and Y. Motome. \textit{Physical Review B}, 103, 024439 (2021).




\bibitem{EuAl4} T. Shang, Y. Xu, D. J. Gawryluk, J. Z. Ma, T. Shiroka, M. Shi, and E. Pomjakushina, \textit{Phys. Rev. B}, 103, L020405 (2021)



\end{thebibliography}
\end{document}